\documentstyle[twocolumn,aps,pra]{revtex}
\draft
\begin{document}
%%%que nos pasa????????????????????
%\documentstyle[preprint,prl,aps]{revtex}
%\documentclass[preprint,prl,aps]{revtex4}
\newcommand{\be}{\begin{equation}}
\newcommand{\ee}{\end{equation}}
\newcommand{\ba}{\begin{eqnarray}}
\newcommand{\ea}{\end{eqnarray}}
\newcommand{\ben}{\begin{eqnarray}}
\newcommand{\een}{\end{eqnarray}}
\newcommand{\nd}{\noindent}
\newcommand{\nl}{\newline}
\newcommand{\nld}{\nl \nd}
% Use the option doublespacing or reviewcopy to obtain double line spacing
% \documentclass[doublespacing]{elsart}
%%%%%%%%%%%%%%%%%%%%+++++++++++++++++++++++++++++++++++++++++++++++++++++++++++++

%%%%%%%%%%%%%%%%%%%%%%%%%%%%%%%%%%%%%%%%%%%%%%%%%%+++++++++++++++++++++++

\title{Fisher information  and thermodynamics'  1st. law}

\author{A. Plastino$^1$, A. R. Plastino$^2$, and
 B. H. Soffer$^3$\\ $^1$ 
Exact Sciences' Faculty-Universidad Nacional de La Plata and
                  National Research Council-CONICET \\           C.C. 727,  1900 La Plata, Argentina \\
$^2$ Department of Physics, University of Pretoria, Pretoria
0002, South Africa \\
$^3$ University of California at Los Angeles, Department
of Electrical Engineering \\ Postal address: 665 Bienveneda
Avenue, Pacific Palisades, California 90272}
\maketitle

\begin{abstract}

\nd It was recently shown, starting from
first principles, that thermodynamics' first law (TFL) can be microscopically obtained without  
need of invoking the adiabatic theorem (AT) [Physica A, {\bf 356}, 167  (2005)]. We show here that a TFL can also 
be found for Fisher's information measure, following a similar procedure. Further, it is proved that enforcing the Fisher-first law requirements in a 
process in which the probability distribution is infinitesimally varied is equivalent to minimizing Fisher's information measure subject to appropriate 
constraints. \newline \newline 
 {\bf Pacs:} 05.30.-d, 05.30.Jp 
%\vskip 2mm  \nd {\bf Keywords:}  First Law, Information theory.

\end{abstract}

%\end{frontmatter}

\section{Introduction}
\label{sec:intro}
\nd The study and revision of some of the principles of statistical mechanics is still an active field 
(see, for instance, among many valuable works, \cite{Lavenda,B01,BC04,TB05,L01,B03,LR03}).  
 In particular, it was shown in \cite{our} that the entire Legendre-transform (LT) of thermodynamics can be microscopically reproduced using Fisher's information measure in place of Boltzmenn's entropy. This abstract LT constitues an essential ingredient that permits one to build up a ``statistical mechanics". Fisher information $I$ allows for such a construction. The desired concavity property of $I$, demonstrated in \cite{our}, further illustrates on its utility as a ``generator" of an underlying microscopic explanation for thermodynamics.

\nd In another, totally disctint  vein, it was also recently shown, starting from first principles and using the point of view of information theory, that thermodynamics' first law (TFL) can be
microscopically obtained without \begin{itemize}
\item 
the need of invoking the adiabatic theorem (AT) with 
\item its awkward dependence on hypothetical external, {\it infinitely slowly} varying
parameters, and also,
\item without needing some other usually invoked constraints and arguments \cite{sandra}. \end{itemize} \nd  We will show here that a ``TFL" can also be found
for Fisher's information measure rather than Shannon's, following a similar procedure.  As a corollary result we demonstrate that, {\it given this Fisher TFL},
using a process in which a sought after unknown probability distribution $f$ is infinitesimally varied is equivalent to seeking $f$ by minimizing Fisher's
information measure subject to appropriate constraints .

\section{Preliminaries}
\nd One of the most important extant information measures was advanced by
R.A.~Fisher in the twenties --for details and discussions we refer
to (for instance)~\cite{roybook,Frieden,Pennini1}--.  Fisher information (FI)
arises as a measure of the expected error in the
measurement of a parameter $\theta$ in a situation governed by a 
family of probability densities 
$\rho(x,\theta)$, where $x$ is, of course, a random variable,
and $\theta $ is the parameter characterizing 
the alluded mono-parameter
family of probability densities~\cite{roybook}. It reads
 
 \be \label{graldefi} I_{General}= \int\,dx\,\rho(x,\theta)\,
 \left[\frac{\partial \ln{\rho}}{\partial \theta}\right]^2.\ee
A particular FI-case of  great
importance~\cite{roybook} is  that of translation families~\cite{roybook,Renyi},
 that is, distribution functions of the  form
  
  \be \label{grasberg}
  \rho(x,\theta) \, = \, \rho(x-\theta).
  \ee
  \noindent  
  In this particular case $\theta$ is a shift parameter.
  In other words, two probability densities with different 
  values of $\theta$ have the same shape, but are displaced with respect to each
  other. It is clear that
  
  \be
  \frac{\partial \rho }{\partial \theta} \, = \, 
  - \,  \frac{\partial \rho }{\partial x},
  \ee    
   \noindent
   and, consequently,
   
   \be \label{iguales} 
   \int\,dx\,\rho(x,\theta)\,
   \left[\frac{\partial \rho}{\partial \theta}\right]^2 \, = \,
   \int\,dx\,\rho(x,\theta)\,
   \left[\frac{\partial \rho}{\partial x}\right]^2.
   \ee    
   From now on, and in order to simplify the notation, the $\theta-$dependence will not be explicitated. 
    Thus, the classical Fisher information
  associated with translations of a, for simplicity's sake, one-dimensional observable $x$
with  probability distribution $f(x)$  becomes~\cite{Hall}

\be \label{define} I=\int \mathrm{d}x\,f(x)\,\left[\frac{\partial \ln
f(x)}{\partial x}\right]^2, \ee and the associated Cramer--Rao
inequality~\cite{roybook,Hall} satisfies the Cramer-Rao relation

\be \Delta x\geq I^{-1}\label{cramer1} \ee where  $\Delta x$ is
the variance of the stochastic variable $x$~\cite{Hall}

\be \Delta x^2=\int \mathrm{d}x\, \rho(x)\, x^2-\left(\int
\mathrm{d}x\, \rho(x)\, x\right)^2, \ee and represents the
mean-square error associated to the above referred to measurement. Notice that, 
unlike the case for Shannon's measure, $I$ is NOT a function of $f(x)$ alone, but also of its derivatives.   

We will derive a relationship involving the internal energy, 
Fisher information and temperature, a generalized pressure and the 
known a priori input constraints of measurement data. Having 
constructed this relationship, it will be most simply interpreted as the desired TFL.

\section{A relation for the differential of Fisher's measure}
%%%%%%%%%%%%%%%%%%%%%%%%%%%%%%%%%%%%%%%%%%%%%%%%%%%%%%%%%%%%%%%%%%%%%%%%%%%%%%%%

\subsection{Notation}
\nd   We start presenting now our new results. We proceed first of all   to  derive a relationship for
$dI$. For this we need to deal with the internal energy
\be \label{interna} U=\langle \hat O_1 \rangle\equiv\langle  H(x) \rangle= Tr
[f H]\equiv \int\,dx\,f(x) H(x), \ee where we have used the abbreviation
\be \label{abrevia} Tr f = \int \,dx\,f(x), \ee
and consider variations $\delta f$ of the distribution function, i.e., a process of the type
\be \label{procesa}  f(x) \rightarrow f(x) + \delta f(x), \ee   
where normalization-preservation  entails that, during the process, \be \label{norm}
Tr[\delta f]=\delta Tr[f]= 0.\ee 
We also have, of course, for other physical quantities $O_i(x)$
\be \label{norm1}
 Tr [\delta f(x) \, O_i(x)]= d \langle O_i(x) \rangle.
\ee

\subsection{The Fisher TFL}
\nd We are going now to rely heavily on the results of Ref. \cite{our}. It was shown there that, if we are discussing a  thermodynamical setting involving
$M$ extensive variables $O_i$, we can recover the formal, Legendre-structure of thermodynamics if we minimize the Fisher information measure $I$ subject to 
appropriate constraints. These are given by the supossedly a-priori known values 

\ben \label{constraints} &  Constraints\,\,in\,\,the\,\,minimization\,\,process\,\, are:
\cr \cr & \langle O_i(x) \rangle=a_i,\,\,(i=1,\ldots,M)) \,\,\,{\rm plus} \cr \cr
& Tr[f]=1, \,\,\,{\rm  normalization \,\,of\,\,} f .\een 
It turns then out that the extremized measure $I$ is a function only of the extensive variables 
\be \label{setting} I=I(\langle O_1(x) \rangle, \,\langle O_2(x) \rangle,\ldots, \langle O_M(x) \rangle), \ee
and that, further \cite{our}, 
\be \label{derivadas} \frac{\partial I}{\partial \langle O_k(x) \rangle}=\lambda_k,\ee
where $\lambda_k,\,\,\,(k=1,\ldots,M)$ are $M$ 
Lagrange multipliers that arise in the course of the minimization procedure.
  From (\ref{setting}) and (\ref{derivadas}) it is now clear that
\be \label{diferIa} dI=\sum_{i=1}^M\, \frac{\partial I}{\partial \langle O_i(x)\rangle}\,d \langle O_i(x)\rangle,\ee
i.e., \be \label{diferI} dI= 
\sum_{i=1}^M\,\lambda_i\,d\langle O_i(x)\rangle.\ee We explicitly separate now the energy term, and {\it define} $\beta=\lambda_{i=1}=1/T$. Here  $T$ will be a ``Fisher-temperature" (FT), whose existence was conjectured in \cite{roybook,Frieden} and proved in \cite{our}. Its properties have been extensively discussed in Ref. \cite{our2}. The FT behaves as $1/T_{orthodox}$, so that an ordinary thermometer, conveniently recalibrated, can measure it. 
Accordingly, a term of the form $TdI$ is the Fisher-equivalent of a ``heat" increment $dQ_F$ \cite{our2} and we are in a position to write  
\be \label{2diferI} dI= \beta dU + \sum_{i=2}^M\,\lambda_i\,d\langle O_i(x)\rangle. \ee 
%\vskip 2mm 
\nd Introducing, finally, the ``generalized pressures" 

\be \label{unamas} p_j=T\lambda_j,\,\,\,(j=2,\ldots,M),\ee (\ref{2diferI}) leads us to

\be \label{ley1} dU= TdI- \sum_{i=2}^M\,p_i\,d\langle O_i(x)\rangle,\ee
which is a ``Fisher's thermodynamics" first-law: note that the variations of the extensive quantities are not independent one of another.
\vskip 2mm 
\nd If we call the differential of work ($dW$), effected at temperature $T$, in the form of \be 
\label{wprk} dW=\sum_{i=2}^M\,p_i\,d\langle O_i(x)\rangle,\ee
and, as stated above, recognize that $TdI$ plays the role of a heat-change  $dQ_F$, as expected (but not explicitly discussed there) from the philosophy espoused in
\cite{our}. Thus,  
we have here shown that work is represented by changes in  expectation values. These
constituted part of our prior knowledge. If a posteriori we
encounter changes, this entails that work has been performed, on or by the system.

\vskip 3mm
\section{TFL as a generator of the probability distribution f}

\nd As a corollary to the above results we infer what would be the case 
if we turn things around and take the TFL (16) as the given, and want 
to find the unknown probability distribution $f$. By considering a change
from $f$ to $f+ \delta f$, with $\delta f$ arbitrary, but constrained so that the TFL (16) is obeyed 
in effecting the probabilities change, we obtain, as a result of that constrained change,
the very same $f$ that is obtained by the method of MaxEnt that 
maximizes Fisher information subject to constraints \cite{our}. 

\nd We now demonstrate this result formaly, in order to make all 
the conditions and steps explicit, as a {\sf THEOREM}:
 
\nd {\sf  HYPOTHESIS: If we constrain the process (\ref{procesa})-(\ref{norm}) forcing it to comply with (\ref{ley1})}, \newline 
\nd {\sf THESIS: this results in a probability distribution that minimizes $I$ subject to the constraints} 

\be \label{restrict} \langle O_i(x) \rangle = a_i,\,\,\,(i=1,\ldots,M),\ee {\sf where the $a_i$ are real numbers supposedly known a-priori}.
 
 \nd {\sf PROOF} \newline 
 The process (\ref{procesa})-(\ref{norm}) generates infinitesimal changes in respectively, $I$, $U$, and the $\langle O_i(x) \rangle,\,\,(i=2,\ldots,M)$,
 and these infinitesimal changes are constrained in the fashion  (\ref{ley1}). 
Thus, for the change $\delta f$ prescribed by (\ref{procesa})-(\ref{norm}) we have (as shown in \cite{our}), starting from  the $I-$definition 
(\ref{define}),
 
\ben \label{Q}  & dI= \int\,dx\,K[f]\, \delta f,\cr\cr & {\rm with\,\,\,a\,\,\,functional-derivative\,\, K=\delta I/\delta f \,\,of\,\,the\,\,form\,\,} \cr  \cr & K=[\frac{d\ln{f}}{dx}]^2+2[\frac{d^2\ln{f}}{dx^2}],\een while the changes 
in the  $\langle O_i(x) \rangle,\,\,(i=2,\ldots,M)$ are governed by (\ref{norm1}), this is, 
\ben \label{otros} & \delta_f [\langle O_m(x)\rangle] =  \delta_f Tr[O_m(x) f(x)]=\cr & = Tr[O_m(x)\delta f],
\,\,(\forall \,\, m). \een  
Of course, $f$ is not yet known. We wish to ``extract" $f$ from the 1st Law-relation (\ref{ley1}).   
We introduce now, for future reference, the  function $W(x)$ defined as
\be \label{considero} W(x)=  \beta H(x) + \sum_{i=2}^M\,\lambda_i\,O_i(x),\ee
where  $\beta$ and the $\lambda$'s are the quantities 
appearing in (\ref{ley1}).  
We are thus demanding simultaneous fulfillment of  (\ref{procesa})-(\ref{norm}), which leads to  
 the two requirements 
\ben \label{duo}  & {\bf (1)}\,\, \int \,dx\,  
[K(x) - \beta H(x)- \sum_{i=2}^M\,\lambda_i\,\langle O_i(x)\rangle]\delta f= \cr\cr & =\int \,dx\, [K(x)-W(x)]\delta f=0 \cr\cr 
 & {\bf (2)}\,\,\int\,dx\, \delta f =0.   \een    
The second one is fullfilled by {\it any arbitrary} variation $\delta f(x)$ such that any possible area under the pertinent curve above the $x-$axis is 
compensated  by another similar one under that axis. 
Now, a {\it special type of variation} among these allow us to obtain an important result from the first relation in (\ref{duo}): consider for the 
associated integral a $\delta f(x)$ that consists of two equal rectangles of vanishing widths and height $C(x)$, centered at points $x_1$ and $x_2$,
respectively, such that one of them points upwards and the other downwards (and thus the contributions of their areas is zero). Since the two points 
$x_1,\,\,\,x_2$ may be located anywhere along the $x-$axis, we are forced to conclude that the rectangle's height, given by
\be \label{rectan} C(x)= [K(x)-W(x)]= constant=-\alpha,\ee which is identical to Eq. (13) of \cite{our}. It is therein shown that, (i)  setting the 
probability distribution $f$ equal to the square of some amplitude $\psi$
\be \label{FPPS} f(x)= \psi(x)^2, \ee
 (ii)  defining the new variable 
\be \label{dos} v(x)=\frac{\partial \ln{\psi(x)}}{\partial x}, \ee 
(iii)  one obtains a Ricatti equation from which one immediately  
gets for the amplitude a Scroedinger-like equation \cite{our}
 
\be \label{erwin} -\frac{\psi''(x)}{2}-\frac{1}{8} \sum_i^M\,\lambda_i\,O_i(x) \psi(x)= \alpha \psi(x).\ee 
Now, (\ref{erwin}) yields simultaneously:
\begin{itemize}
\item the probability distribution that minimizes $I$ subject to the constraints posed by (i) normalization, with Lagrange multiplier $\alpha$ 
and (ii) (\ref{restrict}), with Lagrange multipliers $\lambda_i$, as shown in \cite{our}
\item as demonstrated here, the probability distribution that guarantees that the process  (\ref{procesa})-(\ref{norm}) is effected in such a manner that
the first law requirements (\ref{ley1}) are satisfied, with $\alpha$ equal to minus the rectangle's height in (\ref{rectan}) and the $\lambda_i$'s equal 
to the ratio of the generalized pressures to $\lambda_1\equiv\beta$, i.e., Eq. (\ref{unamas}).
\end{itemize}
{\sf The theorem is proved.}

\vskip 3mm
\section{Discussion and Conclusions}

Two new results have been advanced here: \begin{itemize}
\item We have shown that, within Fisher information theory context, one can derive thermodynamicsÕ first law for the Fisher Thermodynamics without appeal to the
adiabatic theorem, or to any explicit dependence on hypothetical external parameters. Thus, we avoid the need to add, to the theoretical description,
putative infinitely slowly varying external parameters so as to obtain the first law. We find work is represented by changes in expectation values. 
These constituted part of our prior knowledge. If a posteriori we encounter changes, this entails that work has been performed, on or by the system. \item 
 As a corollary we demonstrated that, given the Fisher TFL, using a process in which a sought after unknown probability distribution $f$ is
infinitesimmally varied is equivelant to seeking $f$ by minimizing 
Fisher's information measure subject to appropriate constraints. Plastino and Curado have shown that similar results would obtain for a large class of additional entropies,
namely, those that depend {\it only} on $f$ \cite{curado}, unlike the Fisher measure treated here, that in addition depends on derivatives of $f$ as well.  All these
entropies with the necessary concavity property, and Legendre-transform structure of thermodynamics, allow the construction of a ``statistical mechanics" \cite{curado}.
 \end{itemize}

\vskip 3mm

\nd {\bf Acknowledment:} The authors are very indebted to Prof. B. R. Frieden for fruitful discussions.

\end{document}